\documentclass[a4paper,12pt,eqno]{article}
\usepackage{theorem}
\usepackage{latexsym,amssymb,amsfonts,amsmath}
\newtheorem{theorem}{Theorem}[section]
\newtheorem{lemma}[theorem]{Lemma}

\newtheorem{proposition}[theorem]{Proposition}

\theoremheaderfont{\scshape}

\newcommand{\ZM}{\mathbb{Z}}

\newcommand{\CM}{\mathbb{C}}

\newcommand{\ket}[1]{|#1\rangle}

\title{{\Large {\bf Quantum walks and elliptic integrals}}}
\author{ Norio Konno \\
{\scriptsize Department of Applied Mathematics, 
Faculty of Engineering, 
Yokohama National University}\\
{\scriptsize Hodogaya, Yokohama 240-8501, Japan}\\
{\scriptsize e-mail: konno@ynu.ac.jp}\\
}
\date{\empty }
\begin{document}
\maketitle

\par\noindent
\begin{small}
\baselineskip=24pt
\par\noindent
{\bf Abstract}. P\'olya showed in his 1921 paper that the generating function of the return probability for a two-dimensional random walk can be written in terms of an elliptic integral. In this paper we present a similar expression for a one-dimensional quantum walk. 
\footnote[0]{
{\it Abbr. title:} Quantum walks and elliptic integrals
}
\end{small}

\setcounter{equation}{0}
\section{Introduction}
\label{intro}
The classical random walk (RW) is one of the most popular models to analyze many problems in various fields. The quantum walk (QW) is expected as a counterpart of the RW in quantum systems. The intensive study on QWs has a short history, about ten years, however the attention to the QW increases in the scientific community, for example, physicists, mathematicians, and computer scientists. Recently a book and lecture notes on the QW were published \cite{VAndraca2008,Konno2008b}. For excellent reviews, see \cite{Kempe2003,Kendon2007}. There are two types of QWs, i.e., discrete- and continuous-time walks. The relation between them was investigated in \cite{Strauch2006,Childs2009}. Ambainis et al. \cite{AmbainisEtAl2001} intensively studied the one-dimensional (1D) discrete-time case by the Fourier analysis and path counting method. Recently it was shown by \cite{CanteroEtAl2009} that how the theory of CMV matrices gives a natural tool to study the 1D discrete-time QW. Concerning the CMV matrix, see \cite{CanteroEtAl2003,Simon2007}, for instance. In this paper, we consider the 1D discrete-time QW which is determined by a $2 \times 2$ unitary matrix. We prove that the generating function of the return probability for the 1D QW can be written in terms of an elliptic integral by a path counting method. A similar expression for a 2D RW was presented by P\'olya in 1921 \cite{Polya1921}. The return probability of a 3D RW  is also given by an elliptic integral and was evaluated by Watson in 1939 \cite{Watson1939}. It is known that elliptic integrals and their inverses, i.e., elliptic functions, come from many branches of mathematics and physics. Our statement on the generating function and the elliptic integral may be the tip of an iceberg. Pursuing the study along this line would be one of the interesting future problems on the QW. 

The reminder of the paper is organized as follows. Section 2 treats the definition of the QW. In Sect. 3, we present our main result (Theorem~\ref{saban2}). Section 4 is devoted to proof of Proposition~\ref{saban}. 
Concerning the analysis on the QW, there are some methods, that is, the Fourier analysis, the path counting, the CMV matrix-based method. In our proof, we use a path counting method as in the classical 2D RW case stated in Sect. 3. Finally, we give a conclusion in Sect. 5. 
 
\section{Model}
\label{sec:1}
In this section we define the discrete-time QW on $\ZM$ considered here, where $\ZM$ is the set of integers. In general, the time evolution of the QW is determined by a $2 \times 2$ unitary matrix: 
\begin{align*}
U =
\left[
\begin{array}{cc}
a & b \\
c & d
\end{array}
\right],
\end{align*}
with $a, b, c, d \in \mathbb C$ and $\CM$ is the set of complex numbers. The QW defined by the Hadamard gate $U=H$ with $a=b=c=-d=1/\sqrt{2}$ is often called the {\it Hadamard walk} and has been extensively investigated in the study of the QW:
\begin{eqnarray*}
H=\frac{1}{\sqrt2}
\left[
\begin{array}{cc}
1 & 1 \\
1 &-1 
\end{array}
\right].
\end{eqnarray*}
In the present paper, we focus on the Hadamard walk. The discrete-time QW is a quantum version of the random walk with additional degree of freedom called chirality. The chirality takes values left and right, and it means the direction of the motion of the walker. At each time step, if the walker has the left chirality, it moves one step to the left, and if it has the right chirality, it moves one step to the right. Let define
\begin{eqnarray*}
\ket{L} = 
\left[
\begin{array}{cc}
1 \\
0  
\end{array}
\right],
\qquad
\ket{R} = 
\left[
\begin{array}{cc}
0 \\
1  
\end{array}
\right],
\end{eqnarray*}
where $L$ and $R$ refer to the left and right chirality state, respectively.  To define the dynamics of our model, we divide $U$ into two matrices:
\begin{eqnarray*}
P =
\left[
\begin{array}{cc}
a & b \\
0 & 0 
\end{array}
\right], 
\quad
Q =
\left[
\begin{array}{cc}
0 & 0 \\
c & d 
\end{array}
\right],
\end{eqnarray*}
with $U=P+Q$. The important point is that $P$ (resp. $Q$) represents that the walker moves to the left (resp. right) at position $x$ at each time step in the following. We let $\Xi_{n} (l,m)$ denote the sum of all paths starting from the origin in the trajectory consisting of $l$ steps left and $m$ steps right at time $n$ with $l+m=n$. For example, we have $\Xi_2 (1,1) = Q P + P Q$ and  
\begin{align}
\Xi_4 (2,2) = Q^2 P^2 + P^2 Q^2 + Q P Q P + P Q P Q + P Q^2 P + Q P^2 Q. 
\label{kaede}
\end{align}
In this paper, we take $\varphi_{\ast} = {}^T [1/\sqrt{2},i/\sqrt{2}]$ as the initial qubit state, where $T$ is the transposed operator. Then the probability distribution of the walk starting from $\varphi_{\ast}$ at the origin is symmetric. The probability that our quantum walker is in position $x$ at time $n$ starting from the origin with $\varphi_{\ast}$ is defined by 
\begin{align*}
P (X_{n} =x) = || \Xi_{n}(l, m) \varphi_{\ast} ||^2,
\end{align*}
where $n=l+m$ and $x=-l+m$. Then the return probability at the origin at time $n$ is given by 
\begin{align*}
p_n (0) = P (X_{n} =0).
\end{align*}
By definition, $p_{2n+1} (0) = 0$ for $n \ge 0$ and it is enough to study only even times $2n$.

\section{Result}
In this section we present our results. By definition of the Hadamard walk, we can directly compute 
\begin{align}
p_{0} (0) &= 1, \quad p_{2} (0) = \frac{1}{2} = 0.5, \quad
p_{4} (0) = p_{6} (0) = \frac{1}{8} = 0.125,  
\nonumber 
\\
p_{8} (0) &= p_{10} (0) = \frac{9}{128} = 0.07031 \ldots, 
\quad
p_{12} (0) = p_{14} (0) = \frac{25}{512} = 0.04882 \ldots, 
\nonumber
\\
p_{16} (0) &= p_{18} (0) = \frac{1225}{32768} = 0.03738 \ldots.
\label{nungyo}
\end{align} 
In general, $p_{2n} (0)$ can be written by Legendre polynomials, $P_n (x)$, as follows. As for special functions, see \cite{AndrewsEtAl1999}.
\begin{proposition}
\label{saban}
$p_{0} (0) = 1$ and 
\begin{equation*}
p_{2n} (0) = \frac{1}{2} \> \left[ \left\{ P_{n-1}(0) \right\}^2 + \left\{ P_{n}(0) \right\}^2 \right] \quad (n \ge 1).
\end{equation*}
\end{proposition}
The proof appears in the next section. By Proposition~\ref{saban}, $P_{2n+1}(0)=0$ and 
\begin{equation}
P_{2n}(0) = \frac{1}{2^{2n}} \> {2n \choose n},
\label{tomo}
\end{equation}
we get 
\begin{equation}
p_{4m}(0)=p_{4m+2}(0)= \frac{1}{2} \{ P_{2m} (0) \}^2 = \frac{1}{2^{4m+1}} \> \> {2m \choose m}^2 \quad (m \ge 1). 
\label{nungyo2}
\end{equation}
So (\ref{nungyo}) can also be obtained by (\ref{nungyo2}). From now on we derive our main result (Theorem~\ref{saban2}) from Proposition~\ref{saban}. We begin with
\begin{align*}
\sum_{n=0}^{\infty} p_{n} (0) z^n = \sum_{n=0}^{\infty} p_{2n} (0) z^{2n} =
\frac{1}{2} 
\left[ 
(1+ z^2) \sum_{n=0}^{\infty} \left\{ P_{n}(0) \right\}^2 z^{2n} + 1
\right].
\end{align*}
Therefore by (\ref{tomo}) we see that
\begin{align*}
\sum_{n=0}^{\infty} \left\{ P_{n}(0) \right\}^2 z^{2n} 
= \sum_{n=0}^{\infty} \left\{ P_{2n}(0) \right\}^2 z^{4n} 
= \sum_{n=0}^{\infty} {2n \choose n}^2 \left( \frac{z^2}{4} \right)^{2n}
= \frac{2}{\pi} K(z^2),
\end{align*}
where $K(k)$ is the complete elliptic integral (see \cite{AndrewsEtAl1999}), i.e., 
\begin{align*}
K(k) = \int_{0}^{\pi/2} \frac{d \theta}{\sqrt{1-k^2 \sin ^2 \theta}} 
= \int_{0}^{1} \frac{d x}{\sqrt{(1-x^2)(1-k^2 x^2)}} 
\qquad (0 \le k < 1).
\end{align*}
Then the generating function of the return probability for the 1D Hadamard walk can be expressed by $K(k)$: 
\begin{theorem}
\label{saban2}
\begin{align*}
\sum_{n=0}^{\infty} p_n (0) z^n = \frac{1+z^2}{\pi} \> K(z^2) + \frac{1}{2}. 
\end{align*}
\end{theorem}
Here we consider the generating function for the classical case. The $d$-dimensional classical (simple) RW whose transition probability from the origin to $x$ is given by $1/2d$ for any $x=(x_1, \ldots, x_d) \in \ZM^d$ with $\sum_{k=1}^d |x_k|=1$. Let $p_{n}^{(c,d)} (0)$ denote the return probability for the RW starting from the origin at time $n$. In the 1D RW case, it is well known that 
\begin{align*}
p_{2n}^{(c,1)} (0) = \frac{1}{2^{2n}} \> {2n \choose n}, \quad  p_{2n+1}^{(c,1)} (0)=0 \quad (n \ge 0).
\end{align*}
So we get
\begin{align*}
\sum_{n=0}^{\infty} p_n^{(c,1)} (0) z^n = \frac{1}{\sqrt{1-z^2}}.
\end{align*}
For the 2D RW, we find that 
\begin{align*}
p_{2n}^{(c,2)} (0) = \left\{ p_{2n}^{(c,1)} (0) \right\}^2 
= \frac{1}{4^{2n}} \> {2n \choose n}^2, \quad  p_{2n+1}^{(c,2)} (0)=0 \quad (n \ge 0).
\end{align*}
Concerning the derivation, see for instance \cite{Durrett2004}. Therefore we have
\begin{align*}
\sum_{n=0}^{\infty} p_n^{(c,2)} (0) z^n = 
\frac{2}{\pi} K(z).
\end{align*}
The result was obtained by P\'olya in 1921 (\cite{Polya1921}, p.160). In addition, the above expression is very similar to that for the 1D Hadamard walk (see Theorem~\ref{saban2}). The probability that the 3D RW returns to its starting point, $F=1 - G^{-1} =0.34053 \ldots$, is given by $K(k)$ in the following (\cite{Spitzer1976}, p.103):
\begin{align*}
G = \frac{1}{\pi^2} \> \int_{- \pi}^{\pi} K \left( \frac{2}{3 - \cos \theta} \right) d \theta.
\end{align*}
We should remark that the definition of the complete elliptic integral in \cite{Spitzer1976} is $(2/\pi) \times K(k)$ in our notation. This integral was first evaluated by Watson in 1939 \cite{Watson1939}: 
\begin{align*}
G 
&= 3(18+12\sqrt{2}-10\sqrt{3}-7\sqrt{6}) \> \{ K(2\sqrt{3}+\sqrt{6}-2\sqrt{2}-3)\}^2 \times (2 / \pi)^2 \\
&= 1.51638 \ldots.
\end{align*}

\section{Proof of Proposition~\ref{saban}}
In this section we will prove Proposition~\ref{saban} by a path counting method. First we introduce the useful matrices to compute $\Xi_n (l,m)$:
\[
R =
\left[
\begin{array}{cc}
c  & d \\
0 & 0 
\end{array}
\right], 
\quad
S =
\left[
\begin{array}{cc}
0 & 0 \\
a & b 
\end{array}
\right],
\]
where $a=b=c=-d=1/\sqrt{2}$. In general, products of the matrices $P, \> Q, \> R$ and $S$ are given in Table \ref{tab:1}.
\begin{table}
\caption{Products of $P, Q, R, S$. For example, $P Q = b R$.}
\label{tab:1}  
\centering
\begin{tabular}{c|cccc}
  & $P$ & $Q$ & $R$ & $S$  \\ \hline
$P$ & $aP$ & $bR$ & $aR$ & $bP$  \\
$Q$ & $cS$ & $dQ$ & $cQ$ & $dS$ \\
$R$ & $cP$ & $dR$ & $cR$ & $dP$ \\
$S$ & $aS$ & $bQ$ & $aQ$ & $bS$ 
\end{tabular}
\end{table}
From this table and (\ref{kaede}), we obtain
\begin{equation*}
\label{konno-eqn:araki}
\Xi_4 (2,2) = bcd P + abc Q + b(ad+bc)R + c(ad+bc)S. 
\end{equation*}
We should note that $P, Q, R$ and $S$ form an orthonormal basis of the vector space of complex $2 \times 2$ matrices with respect to the trace inner product $\langle A | B \rangle = $ tr$(A^{\ast}B)$, where $\ast$ means the adjoint operator. So $\Xi_n (l,m)$ has the following form:
\begin{eqnarray*}
\Xi_n (l,m) = p_n (l,m) P + q_n (l,m) Q + r_n (l,m) R + s_n (l,m) S.
\label{konno-eqn:yukitan}
\end{eqnarray*}
The explicit forms of $p_n (l,m), q_n (l,m), r_n (l,m)$ and $s_n (l,m)$ can be computed as follows (see for instance \cite{Konno2002,Konno2005a}):
\begin{lemma}
\label{pro1}
When $l \wedge m (:= \min \{ l, m \}) \ge 1$, we obtain 
\begin{align*}
p_n (l,m) &= \left( \frac{1}{\sqrt{2}} \right)^{n-1} \> \sum_{\gamma =1}^{(l-1) \wedge m} (-1)^{m- \gamma} {l-1 \choose \gamma} {m-1 \choose \gamma -1}, \\
q_n (l,m) &= \left( \frac{1}{\sqrt{2}} \right)^{n-1} \> \sum_{\gamma =1}^{l \wedge (m-1)} (-1)^{m- \gamma -1} {l-1 \choose \gamma -1} {m-1 \choose \gamma}, \\
r_n (l,m) &= s_n (l,m) =\left( \frac{1}{\sqrt{2}} \right)^{n-1} \> \sum_{\gamma =1}^{l \wedge m} (-1)^{m- \gamma} {l-1 \choose \gamma -1} {m-1 \choose \gamma -1}.
\end{align*}
\end{lemma}
By using Lemma~\ref{pro1}, we have
\begin{align*}
\Xi_{2n} (n,n) \varphi_{\ast} 
&= 
\left( \frac{1}{\sqrt{2}} \right)^{2n} \left(-1 \right)^n
\sum_{\gamma =1} ^{n}
\left(-1 \right)^{\gamma}
{n-1 \choose \gamma- 1}^2  
\left[
\begin{array}{cc}
\frac{n}{\gamma} & \frac{n}{\gamma}-2 \\
-\frac{n}{\gamma}+2 & \frac{n}{\gamma} 
\end{array}
\right]
\frac{1}{\sqrt{2}}
\left[
\begin{array}{c}
1  \\
i
\end{array}
\right]
\\
&=
\left( \frac{1}{\sqrt{2}} \right)^{2n+1} \left(-1 \right)^n
\sum_{\gamma =1} ^{n}
\frac{(-1)^{\gamma}}{\gamma}
{n-1 \choose \gamma- 1}^2  
\left[
\begin{array}{c}
n + (n - 2 \gamma) i  \\
-(n - 2 \gamma) + ni
\end{array}
\right].
\end{align*}
Since $p_{2n}(0) = || \Xi_{2n} (n,n) \varphi_{\ast} ||^2$, we get
\begin{align*}
p_{2n}(0)
&=
\left( \frac{1}{2} \right)^{2n}
\left[
\left\{
\sum_{\gamma =1} ^{n}
\frac{(-1)^{\gamma}}{\gamma}
{n-1 \choose \gamma- 1}^2  
n
\right\}^2
+
\left\{
\sum_{\gamma =1} ^{n}
\frac{(-1)^{\gamma}}{\gamma}
{n-1 \choose \gamma- 1}^2  
(n - 2 \gamma)
\right\}^2
\right]
\\
&=
\left( \frac{1}{2} \right)^{2n}
\left[
2 n^2
\left\{
\sum_{\gamma =1} ^{n}
\frac{(-1)^{\gamma}}{\gamma}
{n-1 \choose \gamma- 1}^2 
\right\}^2
\right.
\\
&
\qquad \qquad \qquad \qquad \qquad 
-4n
\sum_{\gamma =1} ^{n}
\sum_{\delta =1} ^{n}
\frac{(-1)^{\gamma+\delta}}{\gamma}
{n-1 \choose \gamma- 1}^2 
{n-1 \choose \delta- 1}^2 
\\
&
\qquad \qquad \qquad \qquad \qquad \qquad 
\left. +4
\sum_{\gamma =1} ^{n}
\sum_{\delta =1} ^{n}
(-1)^{\gamma+\delta}
{n-1 \choose \gamma- 1}^2 
{n-1 \choose \delta- 1}^2 
\right].
\end{align*}
Furthermore we will rewrite $p_{2n}(0)$ by using the Jacobi polynomial, $P^{\nu, \mu} _n (x)$, which is orthogonal on $[-1,1]$ with respect to $(1-x)^{\nu}(1+x)^{\mu}$ with $\nu, \mu > -1$. Then the following relation holds:
\begin{eqnarray}
P^{\nu, \mu} _n (x) = \frac{\Gamma (n + \nu + 1)}{\Gamma (n+1) \Gamma (\nu +1)} \> {}_2F_1(- n, n + \nu + \mu +1; \nu +1 ;(1-x)/2),
\label{konno-eqn:ken}
\end{eqnarray}
where $\Gamma (z)$ is the gamma function. Therefore we see that
\begin{align}
\sum_{\gamma =1} ^{n}
\frac{(-1)^{\gamma -1}}{\gamma}
{n-1 \choose \gamma- 1}^2
&=
{}_2F_1(-(n-1), -(n-1); 2 ; -1)
\nonumber \\
&=
2^{n-1} {}_2F_1(-(n-1), n+1; 2 ; 1/2)
\nonumber \\
&=
\frac{2^{n-1}}{n} P^{(1,0)} _{n-1}(0).
\label{yukari1}
\end{align}
The first equality is given by the definition of the hypergeometric series. The second equality comes from the following relation:
\[
{}_2F_1(a, b; c ;z) = (1-z)^{-a} {}_2F_1(a, c-b; c ;z/(z-1)).
\]
The last equality follows from (\ref{konno-eqn:ken}). In a similar way, we find that 
\begin{align}
\sum_{\gamma =1} ^{n}
\left( -1 \right)^{\gamma -1}
{n-1 \choose \gamma- 1}^2   
= 2^{n-1} P^{(0,0)} _{n-1}(0).
\label{yukari2}
\end{align}
By using (\ref{yukari1}) and (\ref{yukari2}), we get
\begin{align}
p_{2n}(0) 
&=
\left( \frac{1}{2} \right)^{2n}
\left[ 2n^2 \times \frac{2^{2(n-1)}}{n^2} \> \left\{ P^{(1,0)} _{n-1}(0) \right\}^2 
\right.
\nonumber
\\
&
\left. \qquad 
-4n \times \frac{2^{2(n-1)}}{n} \> P^{(1,0)} _{n-1}(0) \> P^{(0,0)} _{n-1}(0) 
+4 \times 2^{2(n-1)} \> \left\{ P^{(0,0)} _{n-1}(0) \right\}^2
\right]
\nonumber 
\\
&= \frac{\left\{P^{(1,0)} _{n-1}(0)\right\}^2}{2} 
- P^{(1,0)} _{n-1}(0) \> P^{(0,0)} _{n-1}(0)
+ \left\{ P^{(0,0)} _{n-1}(0) \right\}^2
\nonumber
\\
&= \frac{1}{2}
\left[ 
\left\{ P^{(1,0)} _{n-1}(0) - P^{(0,0)} _{n-1}(0) \right\}^2 + \left\{P^{(0,0)} _{n-1}(0) \right\}^2 
\right].
\label{imouto1}
\end{align}
From (6.4.20) in \cite{AndrewsEtAl1999}:
\begin{align*}
(n + \alpha +1) P^{(\alpha,\beta)} _{n}(x) - (n+1) P^{(\alpha,\beta)} _{n+1}(x)
= \frac{(2n + \alpha + \beta +2)(1-x)}{2} P^{(\alpha +1,\beta)} _{n}(x),
\end{align*}
we have
\begin{align}
P^{(0,0)} _{n}(0) - P^{(0,0)} _{n+1}(0) = P^{(1,0)} _{n}(0).
\label{imouto2}
\end{align}
Combining (\ref{imouto1}) with (\ref{imouto2}) yields 
\begin{align*}
p_{2n}(0) = 
\left\{P^{(0,0)}_{n}(0) \right\}^2 + \left\{P^{(0,0)}_{n-1}(0) \right\}^2 .
\end{align*}
Since $P_n (0) = P^{(0,0)} _{n}(0)$, the proof of Proposition~\ref{saban} is complete. 

\section{Conclusion}
In the present paper, we showed that the generating function of the return probability for the 1D Hadamard walk can be written in terms of an elliptic integral. Our expression corresponds to the result for the classical 2D RW given by P\'{o}lya in 1921. Indeed, his expression is also written in terms of an elliptic integral. 
One of the future interesting problems is to find similar expressions for general 1D and higher dimensional QWs. 
Resent study on the relation between discrete- and continuous-time QWs, e.g. \cite{Strauch2006,Childs2009}, would allow us to give a new aspect between our result and the corresponding one for the continuous-time case. 
\par
\par
\
\par\noindent
{\bf Acknowledgments.} The author thanks F. Alberto Gr\"unbaum for valuable comments. This work was supported by the Grant-in-Aid for Scientific Research (C) of Japan Society for the Promotion of Science (Grant No. 21540118).

\begin{small}
\bibliographystyle{jplain}

\end{small}

\end{document}